\documentclass[12pt]{iopart}
\usepackage{graphicx}
\usepackage{iopams}
\usepackage{dcolumn}
\usepackage{bm}
\usepackage{amsthm}
\usepackage{amssymb}
\usepackage[flushleft]{caption}
\usepackage{multirow}
\usepackage{subfigure}
\begin{document}

\title{Quantum Fisher information and spin squeezing in one-axis twisting model}

\author{Wei Zhong$^1$, Jing Liu$^1$, Jian Ma$^2$ and Xiaoguang Wang$^1$}

\address{$^1$ Zhejiang Institute of Modern Physics, Department of Physics, Zhejiang University, Hangzhou 310027, China.}
\address{$^2$ Department of Chemistry, Massachusetts Institute of Technology, Cambridge, Massachusetts 02139, USA.}
\ead{xgwang@zimp.zju.edu.cn}

\begin{abstract}
We consider the quantum Fisher information and spin squeezing in one-axis twisting model with a coherent spin state $\vert\theta_{0},\phi_{0}\rangle$. We analytically discuss the dependence of the two parameters: spin squeezing parameter $\xi^2_{K}$ and the average parameter estimation precision $\chi^2$ on the polar angle $\theta_{0}$ and the azimuth angle $\phi_0$. Moreover, we discuss the effects of the collisional dephasing on the dynamics of the two parameters. In this case, the analytical solution of $\xi^2_K$ is also obtained.
\end{abstract}

\submitto{\JPA}
\maketitle

\section{Introduction}
Fisher information first proposed by Fisher \cite{Fisher1925} is a key notion in statistical inference. It quantifies the information that one can extract about a parameter from the observed probability distribution. Moving into the quantum regime, the extension of Fisher information is known as the quantum Fisher information (QFI), which predicts the theoretical achievable limit on the measurement precision in quantum metrology \cite{Helstrom1976,Holevo1982,Braunstein1994,Braunstein1996}.
Besides the application in quantum metrology, QFI also plays an important role in the field of quantum information theory.  In \cite{Luis2010}, Rivas \emph{et al.} gave a nonclassicality criterion derived from the QFI for linear detection scheme. In \cite{Lu2010}, Lu \emph{et al.} proposed the QFI flow as a quantitative measure of the information flow to observe the non-Markovian behavior in open quantum systems.

Additionally, Pezze \emph{et al.} believed that the QFI must have a close relationship with multipartite entanglement \cite{Pezze2009}. They proposed a sufficient condition for particle entanglement as
\begin{equation}
  \chi^2\equiv\frac{N}{\mathcal{F}_{\varphi}}<1,\label{eq:entangement_critierion}
\end{equation}
where $\mathcal{F}_{\varphi}$ denotes the QFI in terms of a parameter $\varphi$ and $N$ is the total number of particles, i.e., atoms and photons. According to quantum Cram\'er-Rao theorem \cite{Helstrom1976,Holevo1982,Braunstein1994,Braunstein1996}, the accuracy of estimation is asymptotically bounded by the inverse of QFI. Thus, the quantity $\chi^2$ in equation (\ref{eq:entangement_critierion}) happens to be the average parameter estimation precision (APEP) of single particles.  Equation (\ref{eq:entangement_critierion}) shows that a quantum state with the APEP-enhancing must be entangled.

Besides entangled states, it has been theoretically and experimentally demonstrated that spin squeezed states (SSSs) can also be used to improve the accuracy of estimation \cite{Pezze2009,Kitagawa&Ueda1993,Wineland1992,Wineland1994,Gross2010,Riedel2010,Santarelli1999,Meyer2001,Leibfried2004,Roos2006,Giovannetti2004,Ma2011Rep}. The concept of SSSs was first established by Kitagawa and Ueda \cite{Kitagawa&Ueda1993}. Meanwhile, they gave the specific schemes to generate these states, such as one-axis twisting (OAT) and two-axis counter-twisting (TAT). It has been demonstrated that the TAT spin squeezing can be transformed from the OAT type by using repeated Rabi pulses \cite{Liu2011}. Moreover, many other mechanisms are promising for producing SSSs \cite{Hald1999,Takeuchi2005,Kuzmich2000,Chaudhury2007,Fernholz2008,Takano2009}. The fact is that the SSSs is one type of quantum entangled states. In \cite{Wang2003}, it has been proved that the SSSs generated via OAT Hamiltonian are pairwise-entangled.

In this paper, we consider the dynamics of the maximal QFI and spin squeezing in the OAT model with the initial state to be taken as a coherent spin state (CSS) $\vert\theta_{0},\phi_{0}\rangle$. We first derive the analytical expressions for the maximal QFI and spin squeezing parameter $\xi^2_K$ \cite{Kitagawa&Ueda1993} in this model. Furthermore, we investigate the dependence of the APEP $\chi^2$ and $\xi^2_K$ on the polar angle $\theta_0$ and  the azimuth angle $\phi_0$ of the initial CSS. In \cite{Jin2009}, It has been found that spin squeezing sensitively depends on $\theta_{0}$ and is independent of $\phi_{0}$. When $\theta_0$ slightly depart from the optimal value of $\pi/2$, the degree of squeezing gets weaker. Similar to $\xi^2_K$, we find that the QFI is also independent of $\phi_0$. But differently, $\chi^2$ for the OAT-induced state is stable during the time evolution and insensitive to $\theta_0$ in the vicinity of $\theta_0=\pi/2$. In addition, decoherence processes are considered in this model. We discuss influences of the collisional dephasing \cite{Vardi2009,Jin2010,Dorner2012,Zhong2013} on the APEP and spin squeezing. In this case, $\xi^2_K$ is analytically solved, while $\chi^2$ is numerically calculated. Our results shows that both the squeezing and the sensitivity are diminished as a result of the collisional dephasing. However, we find that, at the expense of an amount of the evolution time, a near-Heisenberg-scaling precision of $\chi^2$ can still be acquired even in the presence of collisional dephasing.

The outline of this paper is arranged as follows. In section II, we give the notations and definitions of the QFI and spin squeezing. In section III, we devote to consider the maximal QFI and spin squeezing in the OAT model and derive the analytical results. Then we make a comparison between the APEP $\chi^2$ and spin squeezing parameter $\xi^2_K$. In section IV, we consider the effects of collisional depahsing on $\chi^2$ and $\xi^2_K$. Finally, we make a conclusion. In the appendix, the calculation of expectations of the spin components are also given.

\section{Quantum Fisher information and spin squeezing}
In this section, we briefly recall two key definitions. We first introduce the definition of the QFI and the formula for calculation of the maximal QFI. Secondly, we review two kinds of spin squeezing parameters.

\subsection{Quantum Fisher information}
Quantum Fisher information places the fundamental limit to the accuracy of estimating an unknown parameter, playing a paramount role in quantum metrology \cite{Braunstein1994}. It is generalized from the classical Fisher information $F_{\varphi}$ in statistical inference. Given a parameterized family of condition probability densities $p\left(\varepsilon|\varphi\right)$ conditioned on parameter $\varphi\in\mathbb{R}$, the classical Fisher information with respect to $\varphi$ is defined as
\begin{equation}
F_{\varphi}:=\int_{\mathbb{R}}\left(\frac{\partial\ln p\left(\varepsilon|\varphi\right)}{\partial\varphi}\right)^{2}p\left(\varepsilon|\varphi\right)d\varepsilon.
\label{classicalFI}
\end{equation}
where $\varepsilon$ denotes the measurement outcomes of an observable random variable $E$. Note that the observable $E$ here is a continuous variable. If it is discrete, the integral in equation (\ref{classicalFI}) is replaced by a summation. Equation (\ref{classicalFI}) shows that the classical Fisher information is expressed as a statistical variance of an $\varepsilon$-dependent estimator defined by the logarithmic derivative of $p\left(\varepsilon|\varphi\right)$.

In the quantum setting, the QFI is defined by
\begin{equation}
\mathcal{F}_{\varphi}={\rm Tr}\left(\rho_{\varphi}L_{\varphi}^{2}\right),
\label{eq:QFI}
\end{equation}
where $L_{\varphi}$ is the symmetric logarithmic derivative (SLD) operator $L_{\varphi}$ defined by the following equation
\begin{equation}
\frac{\partial}{\partial\varphi}\rho_{\varphi}=\frac{1}{2}\left(L_{\varphi}\rho_{\varphi}+\rho_{\varphi}L_{\varphi}\right).
\label{eq:sld}
\end{equation}
From the above equation, we have $L_{\varphi}^{\dagger}=L_{\varphi}$ and ${\rm Tr}(\rho_{\varphi}L_{\varphi})=0$. Thus the QFI of equation~(\ref{eq:QFI}) can be known as the variance of $L_{\varphi}$ on the state $\rho_{\varphi}$.

In the most fundamental parameter estimation task, the parameter $\varphi$ to be estimated may be generated via some unitary \cite{Lloyd2006} or non-unitary dynamics \cite{Escher2011,DDobrzanski2012}. In this paper, we only focus on the unitary case in which the parametrization process is expressed as
\begin{equation}
  \rho_{\varphi}=\exp\left(-i\varphi J_{\bm n}\right)\rho_{\rm in}\exp\left(i\varphi J_{\bm n}\right),
\label{eq:probe}
\end{equation}
where $J_{\bm n}$ denotes the collective angular momentum in $\bm{n}$ direction with
\begin{equation}
  J_{\bm n}={\bm J}\cdot{\bm n}=\sum_{\alpha=x,y,z}J_{\alpha}n_{\alpha}.
\end{equation}
The components of the collective angular momentum are given by
\begin{equation}
   J_{\alpha}=\sum_{k=1}^{N}\frac{\sigma_{k\alpha}}{2},
   \label{eq:J_collective_operator}
\end{equation}
where $\sigma_{k \alpha}$ denotes the Pauli matrix acting on the $k$th particle. Such unitary process of equation (\ref{eq:probe}) may be used to model the two-mode optical interferometry and the general Ramsey interferometry \cite{Pezze2009,Yurke1986}. With equation (\ref{eq:probe}), the QFI with respect of $\varphi$ can be expressed in the following form \cite{Ma2011Rep,Sun2010,Ma2011,Huang2012,Tan2013}
\begin{equation}
  \mathcal{F}_{\varphi}={\bm n}{\bm C}{\bm n}^{\rm T},
  \label{eq:QFI_mix}
\end{equation}
where $\bm{C}$ is a symmetry matrix with the matrix elements given by
\begin{equation}
  C_{\alpha,\beta} = \sum_{i\neq j}\frac{(p_i-p_j)^2}{p_i+p_j}|\langle i|J_{\alpha}\vert j\rangle\langle i|J_{\beta}\vert j\rangle+\langle i|J_{\beta}\vert j\rangle\langle i|J_{\alpha}\vert j\rangle|^2,
  \label{eq:C_mix}
\end{equation}
where $\lambda_{i}$ and $\left|i\right\rangle $ are the eigenvalues and eigenvectors of $\rho_{\rm in}$, respectively. For pure states, i.e., $\rho_{\varphi}^{2}=\rho_{\varphi}$, the QFI is simply expressed as
\begin{equation}
 \mathcal{F}_{\varphi}=4\Delta J_{{\bm n}}^{2}.
 \label{eq:QFI_pure}
\end{equation}
Corresponding to equation~(\ref{eq:C_mix}), the matrix elements of $\bm{C}$ can be written in the covariance form
\begin{equation}
  C_{\alpha,\beta}={\rm Cov}\left\langle J_{\alpha},J_{\beta}\right\rangle \equiv\frac{1}{2}\left[\langle J_{\alpha}J_{\beta}\rangle+\langle J_{\beta}J_{\alpha}\rangle\right]-\langle J_{\alpha}\rangle\langle J_{\beta}\rangle.
  \label{eq:C_pure}
\end{equation}

To obtain the maximal QFI, we rewrite the variance as
\begin{equation}
  \Delta J_{{\bm n}}^{2}=\bm{n}\bm{O}(\bm{O}^{\rm T}\bm{C}\bm{O})\bm{O}^{\rm T}\bm{n}^{\rm T}=\tilde{\bm{n}}\bm{C}_d\tilde{\bm{n}}^{\rm T},\label{eq:rotation_Jn}
\end{equation}
where $\bm{O}$ is an orthogonal matrix, $\tilde{\bm{n}}$ denotes the rotated direction as $\tilde{\bm{n}}=\bm{n}\bm{O}$, and $\bm{C}_d$ is the diagonal form of $\bm{C}$,
\begin{equation}
  \bm{C}_{d}=\bm{O}^{\rm T}\bm{C}\bm{O}={\rm diag}\{\lambda_i,\lambda_2,\lambda_3\}.
\end{equation}
where $\lambda_i$ are the eigenvalues of $\bm{C}$. Furthermore, equation (\ref{eq:rotation_Jn}) can be expressed as
\begin{equation}
  \max(\Delta J_{{\bm n}}^{2})=\max(\lambda_1 \tilde{n}_1^2+\lambda_2 \tilde{n}_2^2+\lambda_3 \tilde{n}_3^2).
\end{equation}
In above equation, the rotated direction is normalized and satisfies the condition $\tilde{n}_1^2+\tilde{n}_2^2+\tilde{n}_3^2=1$. Assuming that $\lambda_{\max}=\lambda_1$ as the maximal eigenvalue, then we obtain
\begin{equation}
   \mathcal{F}_{\varphi,\max}=4\lambda_{\max},
\end{equation}
with $\tilde{\bm{n}}=(1,0,0)$ and the original direction $\bm{n}=\tilde{\bm{n}}\bm{O}^{\rm T}$.

According to quantum Cram\'er-Rao theorem \cite{Helstrom1976,Holevo1982,Braunstein1994,Braunstein1996}, for $\upsilon$ repetitions of an experiment and any locally unbiased estimator $\hat{\varphi}$ the sensitivity of estimation is bounded by the following inequality
\begin{equation}
\Delta\hat{\varphi}\geq\Delta\varphi_{\rm QCRB}=\frac{1}{\sqrt{\upsilon\mathcal{F}_{\varphi}}}.
\label{eq:QCRB}
\end{equation}
where the equality is saturated in the asymptotic limit $\upsilon\rightarrow\infty$ and $\Delta\varphi_{\rm QCRB}$ is the so-called quantum Cram\'er-Rao bound (QCRB). With equation (\ref{eq:entangement_critierion}), equation (\ref{eq:QCRB}) can be rewritten as
\begin{equation}
\Delta\varphi_{\rm QCRB}=\frac{\chi}{\sqrt{\upsilon N}}.
\label{eq:CRB}
\end{equation}
It shows that when $\chi^{2}=1$ which means that the state is no-entangled, then it can provide a standard-quantum-scaling limit (SQL) on the measurement precision. When $\chi^{2}=1/N$ indicating the maximally entangled states, it gives a Heisenberg-scaling limit (HL) \cite{Pezze2009,Lloyd2006}.

\subsection{Spin squeezing}

Below, we introduce two popular spin squeezing parameters: $\xi_{K}^{2}$ proposed by Kitagawa and Ueda  in analogy to photon squeezing \cite{Kitagawa&Ueda1993}, and $\xi_{W}^{2}$ by Wineland in Ramsey experiments \cite{Wineland1994}. Besides, there are various definitions of spin squeezing which were introduced for certain considerations, and one can refer to reference \cite{Ma2011Rep} for detailed discussion.

The two spin squeezing parameters are defined as follows \cite{Kitagawa&Ueda1993,Wineland1994}
\begin{equation}
\xi_{K}^{2}=\frac{2\min(\Delta J_{{\bm n}_{\bot}}^{2})}{N},
~\xi_{W}^{2}=\frac{N\min(\Delta J_{{\bm n}_{\bot}}^{2})}{\left\vert \left\langle J_{{\rm n}}\right\rangle \right\vert ^{2}},
\label{eq:spin_squeezing_parameters}
\end{equation}
where the subscript ${\bm n}_{\bot}$ denotes the axis perpendicular to the mean spin which is given by $\langle {\bm J}\rangle =\left(\left\langle J_{x}\right\rangle ,\left\langle J_{y}\right\rangle ,\left\langle J_{z}\right\rangle \right)$. In what follows, we use $V_{\pm}$ to denote the maximal and minimal variance of $J_{{\bm n}_{\bot}}$, namely, $V_{+}=\max (J_{{\bm n}_{\bot}})$ and $V_{-}=\min (J_{{\bm n}_{\bot}})$. Evidently, the above two parameters satisfy the following relations
\begin{equation}
  \xi_{W}^{2}=\frac{N^{2}}{2\left\vert \left\langle J_{{\bm n}}\right\rangle \right\vert ^{2}}\,\xi_{K}^{2},
  \,\,{\rm and}\,\,
  \xi_{W}^{2}<\xi_{K}^{2}.
\end{equation}
According to the definition of spin squeezing, when a state satisfies the inequality $\xi_{i}^{2}<1~(i=K,W)$, then it is the SSS. It means that for SSS, its fluctuation of the collective angular momentum is squeezed in one direction and inflated in the another direction, and the fluctuation in the squeezed direction is smaller than the sub-short limit.  It is believed that spin squeezing which is aroused from quantum correlation effect among individual particles, is closely related to multipartite entanglement \cite{Kitagawa&Ueda1993,Wineland1994,Leibfried2004,Soensen&Duan2001}. One of the main applications of the SSSs is used to reduce quantum noise and increase the signal-to-noise ratio in spectroscopy \cite{Wineland1992,Wineland1994,Petrov2007,Chaudhury2006}. Below, we only consider the parameter $\xi^2_{K}$ to characterize spin squeezing.

\section{The maximal QFI and spin squeezing in the OAT model}
In this section, we discuss the maximal QFI and the spin squeezing in the OAT model with a CSS $\vert\theta_{0},\phi_{0}\rangle$ in the absence of noise. First, we introduce the OAT model, then derive the analytical solutions of $\chi^2$ and $\xi_{K}^2$ and finally we make a comparison between $\chi^2$ and $\xi_{K}^2$.

\subsection{One-axis twisting model}
The Hamiltonian of the OAT model is represented by
\begin{equation}
  H=\kappa J_{z}^{2},\label{eq:OAT}
\end{equation}
where $\kappa$ denotes constant number related to the specific systems \cite{Kitagawa&Ueda1993}. This model has been experimentally realized in both two-mode Bose-Einstein condensates (BECs) \cite{Soensen&Duan2001} and light-ensemble interaction in optical cavity \cite{Takeuchi2005,Vuletic2010}. The corresponding time evolution operator to equation (\ref{eq:OAT}) reads
\begin{equation}
  U\left(t\right)=\exp\left(-iHt\right)=\exp\left(-i\kappa J^{2}_{z}t\right).
  \label{eq:time_evolution_operator}
\end{equation}

Generally, the system is initially prepared in a CSS as
\begin{equation}
  \left\vert \theta_{0},\phi_{0}\right\rangle \equiv\left\vert \theta_{0},\phi_{0}\right\rangle _{k}^{\otimes N}=\left(\cos\frac{\theta_{0}}{2}\left\vert \uparrow\right\rangle _{k}+e^{i\phi_{0}}\sin\frac{\theta_{0}}{2}\left\vert \downarrow\right\rangle _{k}\right)^{\otimes N},
  \label{eq:CSS1}
\end{equation}
for $k=1,2,...,N$, which represents as a product state with a set of $N$ elementary spins all pointing in the same direction $\left(\theta_{0},\phi_{0}\right)$. In the above equation, $\theta_{0}$ and $\phi_{0}$ denote the polar and azimuth angles of the polarization of the spins, respectively. By expanding in the basis of the $J_{z}$ operator, the CSS of equation (\ref{eq:CSS1}) can be re-expressed as
\begin{equation}
  \left\vert \theta_{0},\phi_{0}\right\rangle 
  =\sum_{m=-j}^{j}c_{m}(0)\left\vert j,m\right\rangle,
  \label{eq:CSS}
\end{equation}
with the amplitudes
\begin{equation}
  c_{m}(0)={\left(\begin{array}{c} 2j\\j-m\end{array}\right)}^{1/2}\left(\sin\frac{\theta_{0}}{2}\right)^{j-m}\left(\cos\frac{\theta_{0}}{2}\right)^{j+m}e^{i\left(j-m\right)\phi_{0}}.
\end{equation}

Under the nonlinear evolution of equation (\ref{eq:time_evolution_operator}), the state of the system at time $t$ becomes
\begin{equation}
  \left\vert \psi\left(t\right)\right\rangle =U\left(t\right)\left\vert \theta_{0},\phi_{0}\right\rangle =\sum_{m=-j}^{j}c_{m}\left(0\right)\exp\left(-i\kappa m^{2}t\right)\left\vert j,m\right\rangle.
\label{eq:OAT_state}
\end{equation}
Below, we will calculate the dynamics of the maximal QFI $\mathcal{F}_{\varphi,\max}$ and $\xi_{K}^2$ for this state. Here we only take account on the calculation of $\mathcal{F}_{\varphi,\max}$. The detailed computation processed of $\xi^2_K$ are given in \cite{Jin2009}.

\subsection{The exact solutions of $\chi^{2}$ and $\xi^{2}_{K}$ in OAT model}

In the Heisenberg picture, with equation (\ref{eq:time_evolution_operator}), the time evolution of the ladder operator $J_{+}$ could be exactly calculated
\begin{equation}
  \tilde{J}_{+}=U(t)^{\dag}J_{+}U(t)=J_{+}\exp\left[i\mu\left(J_{z}+\frac{1}{2}\right)\right],
\end{equation}
where the superscript tilde signs that the operator is time-dependent and $\mu=2\kappa t$. Then we can analytically derive a set of expectations $\langle \tilde{J}_{+}\rangle ,\langle \tilde{J}_{+}^{2}\rangle$ and $\langle \tilde{J}_{+}(\tilde{J}_{z}+\frac{1}{2})\rangle $ for the OAT-induced state of equation (\ref{eq:OAT_state}). The detailed derivation is presented in the appendix. With equation (\ref{eq:a1}), one can easily obtain the following equations
\begin{equation}
   \langle\tilde{J}_{x}\rangle={\rm Re}\langle\tilde{J}_{+}\rangle,
   \,{\rm and}\,
   \langle\tilde{J}_{y}\rangle={\rm Im}\langle\tilde{J}_{+}\rangle.\label{eq:JxJy}
\end{equation}
Considering the commute relations $[H,J^{2}]=[H,J_{z} ]=0$, one readily derive
\begin{eqnarray}
  \langle\tilde{J}_{z}\rangle
  &=& \langle J_{z}\rangle = j\cos\theta_{0}, \label{eq:Jz}\\
  \langle \tilde{J}_{z}^{2}\rangle
  &=& \langle J_{z}^{2}\rangle
    = \frac{j}{2}+j\left(j-\frac{1}{2}\right)\cos^{2}\theta_{0}, \label{eq:J2z}\\
  \langle \tilde{\bm{J}}^{2}\rangle
  &=&  j(j+1).\label{eq:J2}
\end{eqnarray}

Now, we choose the new orthogonal vectors as
\begin{eqnarray}
\bm{n}_{1} & = & \left(-\sin\phi,\cos\phi,0\right),\\
\bm{n}_{2} & = & \left(-\cos\theta\cos\phi,-\cos\theta\sin\phi,\sin\theta\right),\\
\bm{n}_{3} & = & \left(\sin\theta\cos\phi,\sin\theta\sin\phi,\cos\theta\right),
\end{eqnarray}
where $\bm{n}_{3}$ is the mean spin direction, $\bm{n}_{1}$
and $\bm{n}_{2}$ are the other two directions perpendicular to
$\bm{n}_{3}$, and the trigonometric functions are resolved by
\begin{equation}
  \cos\theta=\langle\tilde{J}_{z}\rangle/\mathcal{R},~\sin\theta=r/\mathcal{R},~{\rm and}~\tan\phi=\langle\tilde{J}_{y}\rangle/\langle\tilde{J}_{x}\rangle,
\end{equation}
with the length of the mean spin
\begin{equation}
  \mathcal{R}=\vert \langle\tilde{\bm{J}}\rangle\vert =\sqrt{\langle\tilde{J}_{x}\rangle^{2}+\langle\tilde{J}_{y}\rangle^{2}+\langle\tilde{J}_{z}\rangle^{2}},
\end{equation}
and
\begin{equation}
  r=\mathcal{R}\sin\theta=\sqrt{\langle\tilde{J}_{x}\rangle^{2}+\langle\tilde{J}_{y}\rangle^{2}}.
\end{equation}
It is easy to verify that $\langle J_{\bm{n}_{1}}\rangle=\langle J_{\bm{n}_{2}}\rangle=0$
and $\langle J_{\bm{n}_{3}}\rangle=\vert \langle\bm{\tilde{J}}\rangle\vert $.

\begin{table*}[t]
\caption{The matrix elements of the symmetry covariance matrix $\bm{C}$ of equation (\ref{eq:matrix_C}) are fully represented by six terms: $\langle \tilde{\bm{J}}^{2}\rangle$,
$\langle\tilde{J}_{z}\rangle$, $\langle\tilde{J}_{z}^{2}\rangle$, $\langle\tilde{J}_{+}\rangle e^{-i\phi}$,
$\langle\tilde{J}_{+}^{2}\rangle e^{-i2\phi}$ and $\langle\tilde{J}_{+}(2\tilde{J}_{z}+1)\rangle e^{-i\phi}$.}
\label{tab:table2}
\centering{}
\begin{tabular}{|c|c|}
\hline
\parbox[c]{3cm}{%
$\langle J_{\bm{n}_{3}}\rangle$%
}  & %
\parbox[c]{11.5cm}{%
$\sin\theta\,{\rm Re}[\langle\tilde{J}_{+}\rangle e^{-i\phi}]+\cos\theta\,\langle\tilde{J}_{z}\rangle$%
}  \\\hline
\parbox[c]{3cm}{%
$\langle J_{\bm{n}_{1}}^{2}\rangle$%
}  & %
\parbox[c]{11.5cm}{%
$\frac{1}{2}\,[\langle\tilde{\bm{J}}^{2}\rangle-\langle\tilde{J}_{z}^{2}\rangle]-\frac{1}{2}\,{\rm Re}[\langle\tilde{J}_{+}^{2}\rangle e^{-i2\phi}]$%
} \\\hline
\parbox[c]{3cm}{%
$\langle J_{\bm{n}_{2}}^{2}\rangle$%
}  & %
\parbox[c]{11.5cm}{%
$\frac{1}{2}\cos^{2}\theta\,[\langle\tilde{\bm{J}}^{2}\rangle-\langle\tilde{J}_{z}^{2}\rangle]+\sin^{2}\,\theta\langle\tilde{J}_{z}^{2}\rangle+\frac{1}{2}\cos^{2}\theta\,{\rm Re}[\langle\tilde{J}_{+}^{2}\rangle e^{-i2\phi}]-\frac{1}{2}\sin2\theta\,{\rm Re}[\langle\tilde{J}_{+}(2\tilde{J}_{z}+1)\rangle e^{-i\phi}]$%
}  \\\hline
\parbox[c]{3cm}{%
$\langle J_{\bm{n}_{3}}^{2}\rangle$%
}  & %
\parbox[c]{11.5cm}{%
$\frac{1}{2}\sin^{2}\theta\,[\langle\tilde{\bm{J}}^{2}\rangle-\langle\tilde{J}_{z}^{2}\rangle]+\cos^{2}\theta\,\langle\tilde{J}_{z}^{2}\rangle+\frac{1}{2}\sin^{2}\theta\,{\rm Re}[\langle\tilde{J}_{+}^{2}\rangle e^{-i2\phi}]+\frac{1}{2}\sin2\theta\,{\rm Re}[\langle\tilde{J}_{+}(2\tilde{J}_{z}+1)\rangle e^{-i\phi}]$%
}  \\\hline
\parbox[c]{3cm}{%
$\langle[J_{\bm{n}_{1}},J_{\bm{n}_{2}}]_{+}\rangle$%
}  & %
\parbox[c]{11.5cm}{%
$-\cos\theta\,{\rm Im}[\langle\tilde{J}_{+}^{2}\rangle e^{-i2\phi}]+\sin\theta\,{\rm Im}[\langle\tilde{J}_{+}(2\tilde{J}_{z}+1)\rangle e^{-i\phi}]$%
} \\\hline
\parbox[c]{3cm}{%
$\langle[J_{\bm{n}_{1}},J_{\bm{n}_{3}}]_{+}\rangle$%
}  & %
\parbox[c]{11.5cm}{%
$\sin\theta\,{\rm Im}[\langle\tilde{J}_{+}^{2}\rangle e^{-i2\phi}]+\sin\theta\,{\rm Im}[\langle\tilde{J}_{+}(2\tilde{J}_{z}+1)\rangle e^{-i\phi}]$%
} \\\hline
\parbox[c]{3cm}{%
$\langle[J_{\bm{n}_{2}},J_{\bm{n}_{3}}]_{+}\rangle$%
}  & %
\parbox[c]{11.5cm}{%
$-\frac{1}{2}\sin2\theta\,[\langle\tilde{\bm{J}}^{2}\rangle-3\langle\tilde{J}_{z}^{2}\rangle+\langle\tilde{J}_{+}^{2}\rangle e^{-i2\phi}]-\cos2\theta\,{\rm Re}[\langle\tilde{J}_{+}(2\tilde{J}_{z}+1)\rangle e^{-i\phi}]$%
}\\
\hline
\end{tabular}
\end{table*}

To calculate $\mathcal{F}_{\varphi,\max}$, we first should determine the symmetry covariance matrix ${\bm C}$ of equation (\ref{eq:C_pure}). In the new orthogonal basis $(\bm{n}_1,\bm{n}_2,\bm{n}_3)$, the matrix $\bm{C}$ is given by
\begin{equation}
  \bm{C}=\left[\begin{array}{ccc}
\langle J_{\bm{n}_{1}}^{2}\rangle & \langle\left[J_{\bm{n}_{1}},J_{\bm{n}_{2}}\right]_{+}\rangle & \langle\left[J_{\bm{n}_{1}},J_{\bm{n}_{3}}\right]_{+}\rangle\\
\langle\left[J_{\bm{n}_{1}},J_{\bm{n}_{2}}\right]_{+}\rangle & \langle J_{\bm{n}_{2}}^{2}\rangle & \langle\left[J_{\bm{n}_{2}},J_{\bm{n}_{3}}\right]_{+}\rangle\\
\langle\left[J_{\bm{n}_{1}},J_{\bm{n}_{3}}\right]_{+}\rangle & \langle\left[J_{\bm{n}_{2}},J_{\bm{n}_{3}}\right]_{+}\rangle & \langle J_{\bm{n}_{3}}^{2}\rangle-\langle J_{\bm{n}_{3}}\rangle^{2}\end{array}\right].
\label{eq:matrix_C}
\end{equation}
where the matrix elements of $\bm{C}$ are fully determined by six terms: $\langle \tilde{\bm{J}}^{2}\rangle$,
$\langle\tilde{J}_{z}\rangle$, $\langle\tilde{J}_{z}^{2}\rangle$, $\langle\tilde{J}_{+}\rangle e^{-i\phi}$,
$\langle\tilde{J}_{+}^{2}\rangle e^{-i2\phi}$ and $\langle\tilde{J}_{+}(2\tilde{J}_{z}+1)\rangle e^{-i\phi}$ (see table \ref{tab:table2}). Diagonalize $\bm{C}$ and find the maximal eigenvalue of $\bm{C}$ as $\lambda_{\max}$. Then we have
\begin{equation}
\mathcal{F}_{\varphi,\max}\left(t\right)=4\,\lambda_{\max}.
\label{eq:excat F}
\end{equation}
It deserves to note that all matrix elements of $\bm{C}$ are independent of $\phi_{0}$. It means that the maximal QFI $\mathcal{F}_{\varphi,\max}(t)$ does not depend on $\phi_{0}$, which is the same as the spin squeezing parameters $\xi_{K}^{2}$ \cite{Jin2009}. This result can be easily understood by considering the commute relation $\left[H,J_{z}\right]=0$ and the unitary invariance property of the QFI which states that the value of the QFI remains invariant under the unitary evolution being independent of $\varphi$ \cite{Helstrom1976,Holevo1982}. Due to the system Hamiltonian commuting with $J_{z}$, we have $\left[R\left(\phi_{0},J_{z}\right),U(t)\right]=0$, where $R\left(\phi_{0},J_{z}\right)$ denotes a rotation around $x$-axis by angle $\phi_{0}$. Then the OAT-induced state of equation (\ref{eq:OAT_state}) satisfies the following chain equalities
\begin{eqnarray}
\left\vert \psi(t)\right\rangle
&=&U(t)\left\vert \theta_{0},\phi_{0}\right\rangle
=U(t)\,R(\phi_{0},J_{z})\left\vert \theta_{0},0\right\rangle =R(\phi_{0},J_{z})\,U(t)\left\vert \theta_{0},0\right\rangle.
\end{eqnarray}
According to the unitary invariance property, the QFI of the state $U(t)\left\vert \theta_{0},\phi_{0}\right\rangle$ equals that of the state $U(t)\left\vert \theta_{0},0\right\rangle$ which is $\phi_{0}$-independent.
Thus it means that the QFI is $\phi_{0}$-independent. In what follows, we set $\phi_{0}=0$ as the initial CSS.

In the above derivation of the maximal QFI, there involves the diagonalization of the matrix $\bm{C}$. It indicates that $\mathcal{F}_{\varphi,\max}$  corresponds to the maximal variance of the angular momentum operator in the coordinate sphere. Now, we only focus on the maximal and minimal variances in the plane perpendicular to the mean spin direction. We introduce a component of the collective angular momentum normal to the mean spin direction as
\begin{equation}
  J_{\bm{n}_{\bot}}
  =e^{-i\vartheta J_{\bm{n}_3}}J_{\bm{n}_1}e^{i\vartheta J_{\bm{n}_3}}
  =J_{\bm{n}_1}\cos\vartheta+J_{\bm{n}_2}\sin\vartheta.
\end{equation}
Due to $\langle J_{\bm{n}_{1}}\rangle=\langle J_{\bm{n}_{2}}\rangle=0$, we have $\langle J_{\bm{n}_{\bot}}\rangle=0$. Then the variance of $J_{\bm{n}_{\bot}}$ is
\begin{equation}
  \langle \Delta J_{\bm{n}_{\bot}}^2\rangle
  =\langle J_{\bm{n}_{\bot}}^2\rangle
  =\frac{1}{2}[\mathcal{C}+\sqrt{\mathcal{A}^2+\mathcal{B}^2}\cos(2\vartheta-2\delta)],
\end{equation}
with $\tan2\delta=\mathcal{B}/\mathcal{A}$, where the coefficients $\mathcal{A}=\langle J_{\bm{n}_{1}}^{2}-J_{\bm{n}_{2}}^{2}\rangle,\mathcal{B}=\langle\left[J_{\bm{n}_{1}},J_{\bm{n}_{2}}\right]_{+}\rangle$
and $\mathcal{C}=\langle J_{\bm{n}_{1}}^{2}+J_{\bm{n}_{2}}^{2}\rangle=j\left(j+1\right)-\langle J_{\bm{n}_{3}}^{2}\rangle$ (see table \ref{tab:table2}).
The maximal and minimal fluctuations are expressed as
\begin{equation}
  V_{\pm}=(\mathcal{C}\pm\sqrt{\mathcal{A}^{2}+\mathcal{B}^{2}})/2,\label{eq:V_pm}
\end{equation}
where $V_{+}$ occurs along the optimal anti-squeezed angle as $\vartheta_{\rm op}=\left[\tan^{-1}\left(\mathcal{B}/\mathcal{A}\right)\right]/2$ and $V_{-}$ along the optimal squeezed angle as $\vartheta_{\rm op}=\left[\tan^{-1}\left(\mathcal{B}/\mathcal{A}\right)+\pi\right]/2$.
It has been numerically found that the maximal QFI for OAT-induced state is \cite{Liu2010}
\begin{equation}
\mathcal{F}_{\varphi,\max}(t)=4\max\{V_{+},\Delta J_{\bm{n}_3}^2\}.
\label{eq:simple F}
\end{equation}
For odd values of $N$, $\mathcal{F}_{\varphi,\max}$ always occurs in the $\bm{n}_1\bm{n}_2$-plane, i.e., $\mathcal{F}_{\varphi,\max}(t)=4V_{+}$. For even $N$ when $\Delta J_{\bm{n}_3}^2=V_{+}$, $\mathcal{F}_{\varphi,\max}$ occurs in the ${\bm{n}_3}$ direction \cite{Liu2010}.

In addition, with $V_{-}$, we recover the analytical expression of $\xi_{K}^{2}$ according to equation (\ref{eq:spin_squeezing_parameters}) given in \cite{Jin2009}. One can refer to \cite{Jin2009} for a detailed discussion on $\xi_{K}^{2}$ in OAT model.

\subsection{A comparison of the parameters between $\chi^{2}$ and $\xi_{K}^{2}$ }

\begin{figure}[t]
\centering
\subfigure{ \includegraphics[width=7.5cm,height=6cm]{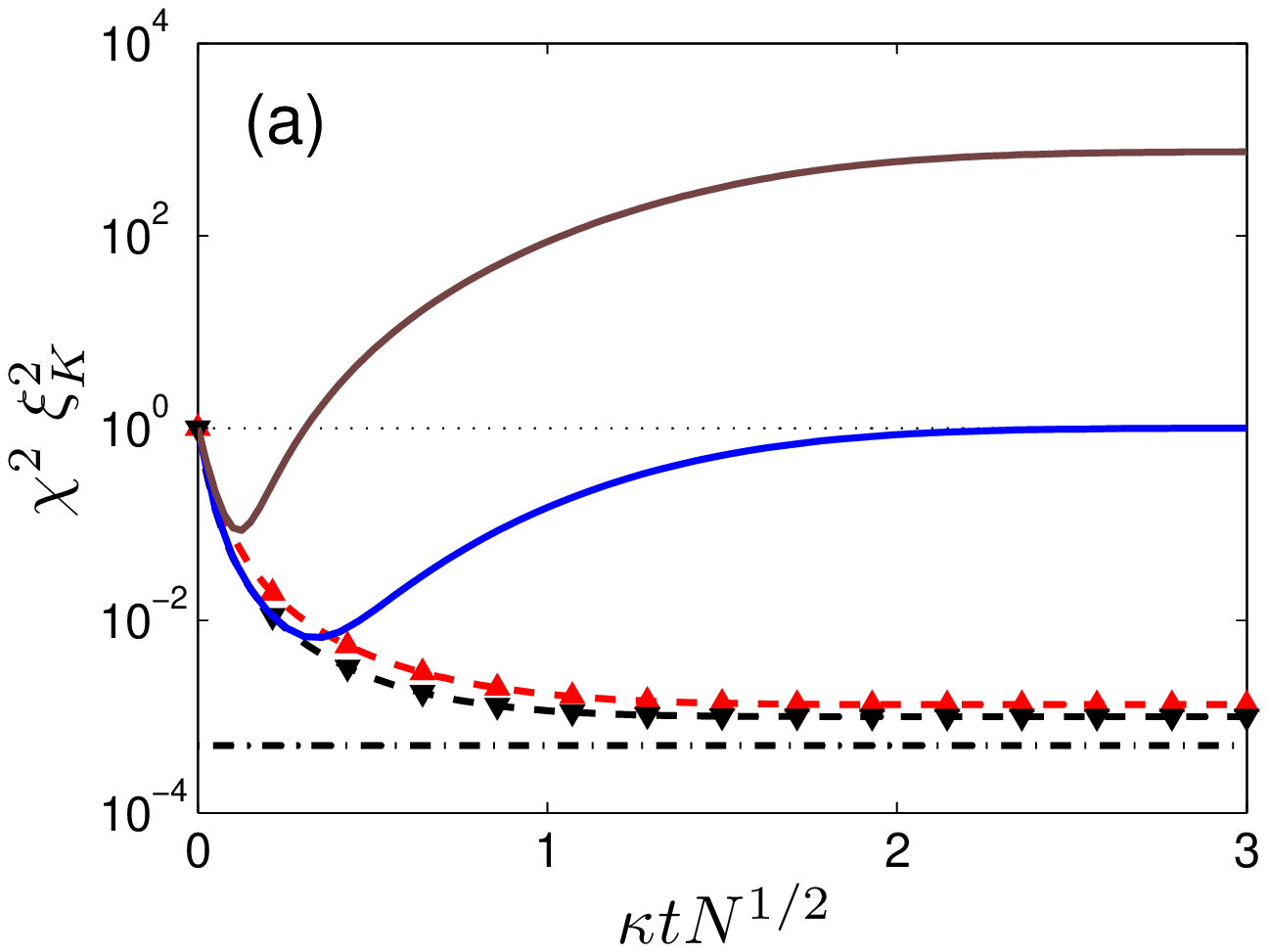}}
\subfigure{ \includegraphics[width=7.5cm,height=6cm]{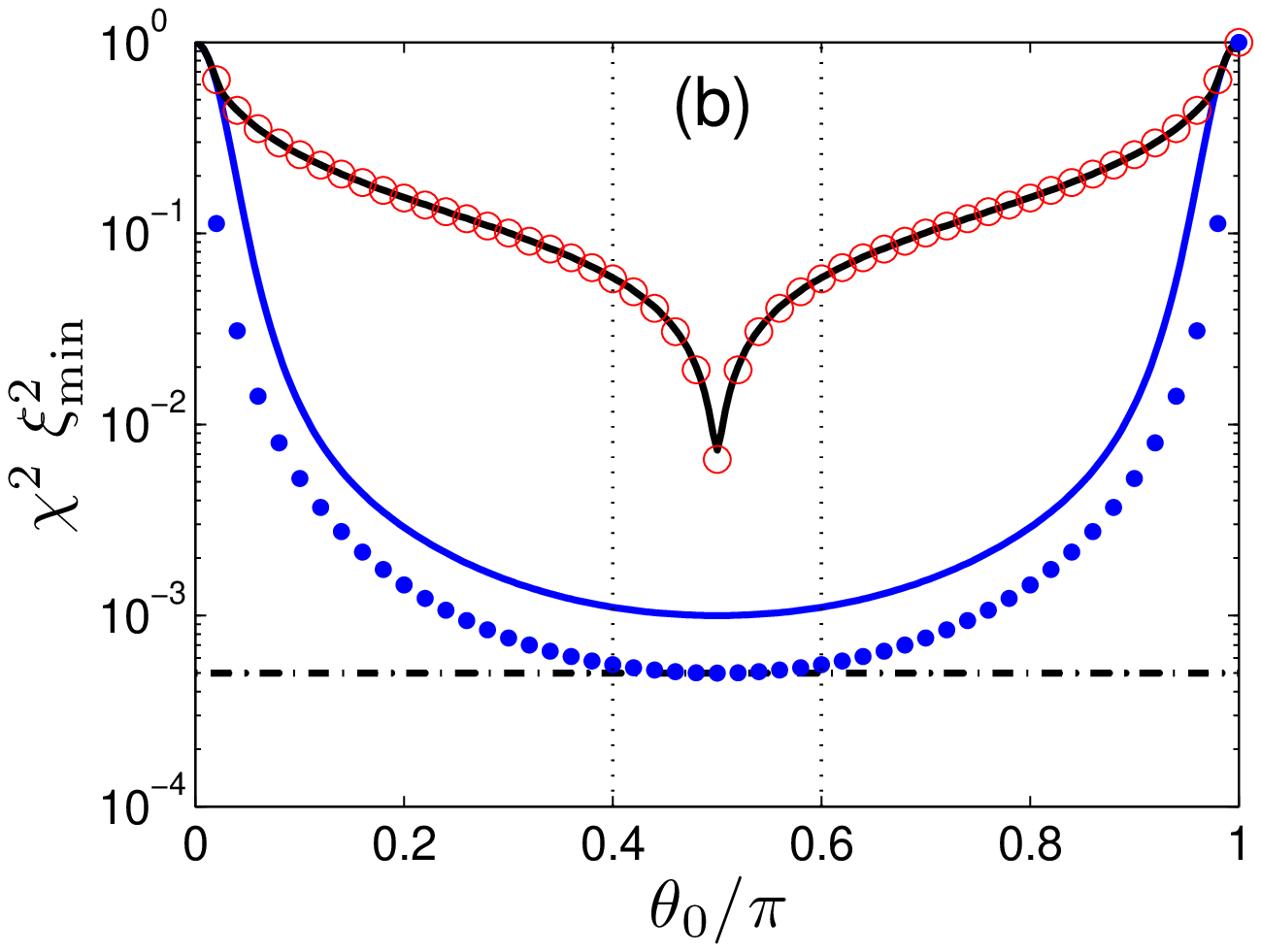}}
\centering
\caption{(Color online) (a) Plots of $\xi_{K}^{2}$ (solid lines) and $\chi^{2}$ (dashed lines) in terms of the re-scaled time $\kappa t\sqrt{N}$ with $N=2\times10^{3}$ for two different initial CSSs: $\vert\pi/2,0\rangle$ (black and blue) and $\vert\pi/3,0\rangle$ (red and brown). Upward triangle and downward triangle are plotted by $\chi^2=N/4V_+$ from equation (\ref{eq:simple F}). (b) Plots of the minimum value of parameters: $\xi_{K,\min}^{2}$ (red circles) and $\chi^{2}_{\min}$ (blue dots) in terms of $\theta_{0}$ for $N=2\times10^{3}$. Black soiled curve denotes $\xi_{W,\min}^{2}$ given by equation (\ref{eq:spin_squeezing_parameters}) \cite{Jin2009}. Solid blue curve denotes the value of $\chi^{2}$ at $\kappa t=3/\sqrt{N}$ (being on the plateau in figure \ref{fig:fig1}(a)). In both (a) and (b), dot-dashed black lines represent the fundamental HL $1/N$.}
\label{fig:fig1}
\end{figure}

Having obtained the analytical solutions of $\chi^2$ and $\xi^2_K$, we make some discussions about the dynamics of the two parameters and the dependence of them on the polar angle $\theta_0$. Figure \ref{fig:fig1}(a) displays the dynamics of $\chi^2$ and $\xi_{K}^{2}$ versus time for the OAT-induced state of equation (\ref{eq:OAT_state}) in terms of two different initial CSSs $\vert\theta_{0}=\pi/2,\phi_{0}=0\rangle$ and $\vert\pi/3,0\rangle$. It shows that $\xi^2_K$ obviously change when $\theta_0$ varies from $\pi/2$ to $\pi/3$. Meanwhile, we see that both the time interval during which squeezing occurs and the degree of squeezing are shrunken. Different from the behavior of $\xi^2_k$, $\chi^{2}$ is rather stable, only having a small amount of reduction when $\theta_0$ changes from $\pi/2$ to $\pi/3$. As is shown in figures \ref{fig:fig1}(a) and \ref{fig:fig2}(b) for a total time period, $\chi^{2}$ decreases quickly from the initial value of $1$ at beginning, then reaches a plateau $(\chi^2\propto2/N)$ in the time interval of $3/\sqrt{N}\lesssim\kappa t\lesssim\pi/2$, and arrives its minimal value at time point of $\kappa t=\pi/2$ \cite{Pezze2009}. For the case of $\theta_0=\pi/2$, the minimal value of $\chi^2$ at $\kappa t=\pi/2$ corresponds to the HL $1/N$. More importantly, when $\xi_{K}^{2}\geq1$, which indicates that the state is no-squeezed, we still have $\chi^{2}<1$, which means this state is entangled and helpful for improvement of APEP.
Since $\chi^2$ stays on the plateau for quite a long time, it means that one can always acquire a near-Heisenberg limit $\propto2/N$ with the OAT-induced state only at the expense of a bit precision.

In figure \ref{fig:fig1}(b), we plot $\chi_{\min}^2$ and $\xi_{\min}^2$ as the function of $\theta_0$. As is shown in \cite{Jin2009}, in short-time limit ($\kappa t\ll 0$) and the large particle number ($N\gg 0$) case, $\xi_{K,\min}^2$ occurs at
\begin{equation}
  \kappa t_{\min}=\frac{3^{1/6}\left(2j\sin^{2}\theta_{0}\right)^{-2/3}}
  {\left(1+9j\sin^{2}\theta_{0}\cos^{2}\theta_{0}\right)^{1/6}}.
\end{equation}
We also plot $\chi^2$ (solid blue curve) at $\kappa t=3/\sqrt{N}$ (being on the plateau in figure \ref{fig:fig1}(b)) versus $\theta_0$. It clearly shows that both the strongest squeezing and the best APEP occur for $\theta_0=\pi/2$. However, when $\theta_{0}$ slightly deviates from the optimal value of $\pi/2$, the degree of squeezing is decreased. Different from the behavior of $\xi_{\min}^2$, $\chi^{2}$ is fairly robust against $\theta_0$ in the vicinity of $\theta_{0}=\pi/2$. We see that $\chi^{2}$ is stable in the regime of $\vert\theta_{0}-\pi/2\vert <2\pi/5$ as marked by vertical dotted lines.

The fact is that the ideal optimal CSS $\vert\pi/2,0\rangle$ is hard to generate in experiment. It is experimentally shown that $98 \%$ of the atoms can be prepared in the ideal state through optical pumping at present \cite{Fernholz2008}. According to $\chi^2$, we find that the accuracy of estimation given by the state nonlinearly evolved from an unideal initial CSS $\theta_{0}\sim\pi/2$ is almost the same as the sensitivity given by the OAT-induced state from the optimal CSS.

\section{The maximal QFI and spin squeezing under collisional dephasing}

Up to now, we have not considered the influences of decoherence on the dynamics of $\chi^2$ and $\xi^2_{K}$ in the OAT model. In the real experiment, the collisional dephasing process always exists due to the interaction between the atoms and thermal reservoir in two component atomic BEC system. Then the time evolution of the system is governed by the following master equation \cite{Jin2010,Vardi2009,Dorner2012,Zhong2013}
\begin{equation}
  \dot{\rho}\left( t\right)
  =i[\rho,\,\kappa J_{z}^{2}]+\mathcal{L}\,\rho,
  \label{eq:master_eq}
\end{equation}
with
\begin{equation}
  \mathcal{L}\,\rho
  \equiv\Gamma(2 J_{z}\rho(t)J_{z}-\rho(t)J^2_{z}-J^2_{z}\rho(t)),
\end{equation}
where $\mathcal{L}$ denotes the Lindblad superoperator, $\Gamma$ denotes the collisional dephasing rate, and $\rho$ denotes the reduced density operator of the system in the interaction picture. From equation (\ref{eq:master_eq}), the time evolutions of the density matrix  elements are given as follows
\begin{eqnarray}
  \rho_{m,n}\left(t\right)
  &\equiv& \left\langle j,m\right|\rho\left(t\right)\left|j,n\right\rangle
  = \rho_{m,n}\left(0\right)\exp[i(n^2-m^2)\tau-(m-n)^{2}\gamma\tau],
  \label{eq:master_eq result}
\end{eqnarray}
where we have set $\tau=\kappa t$ and $\gamma=\Gamma/\kappa$. It shows that dephasing makes the diagonal elements of the density matrix unchanged and energy conserved.

\begin{figure}[t]
\centering
\subfigure{ \includegraphics[width=7.5cm,height=6cm]{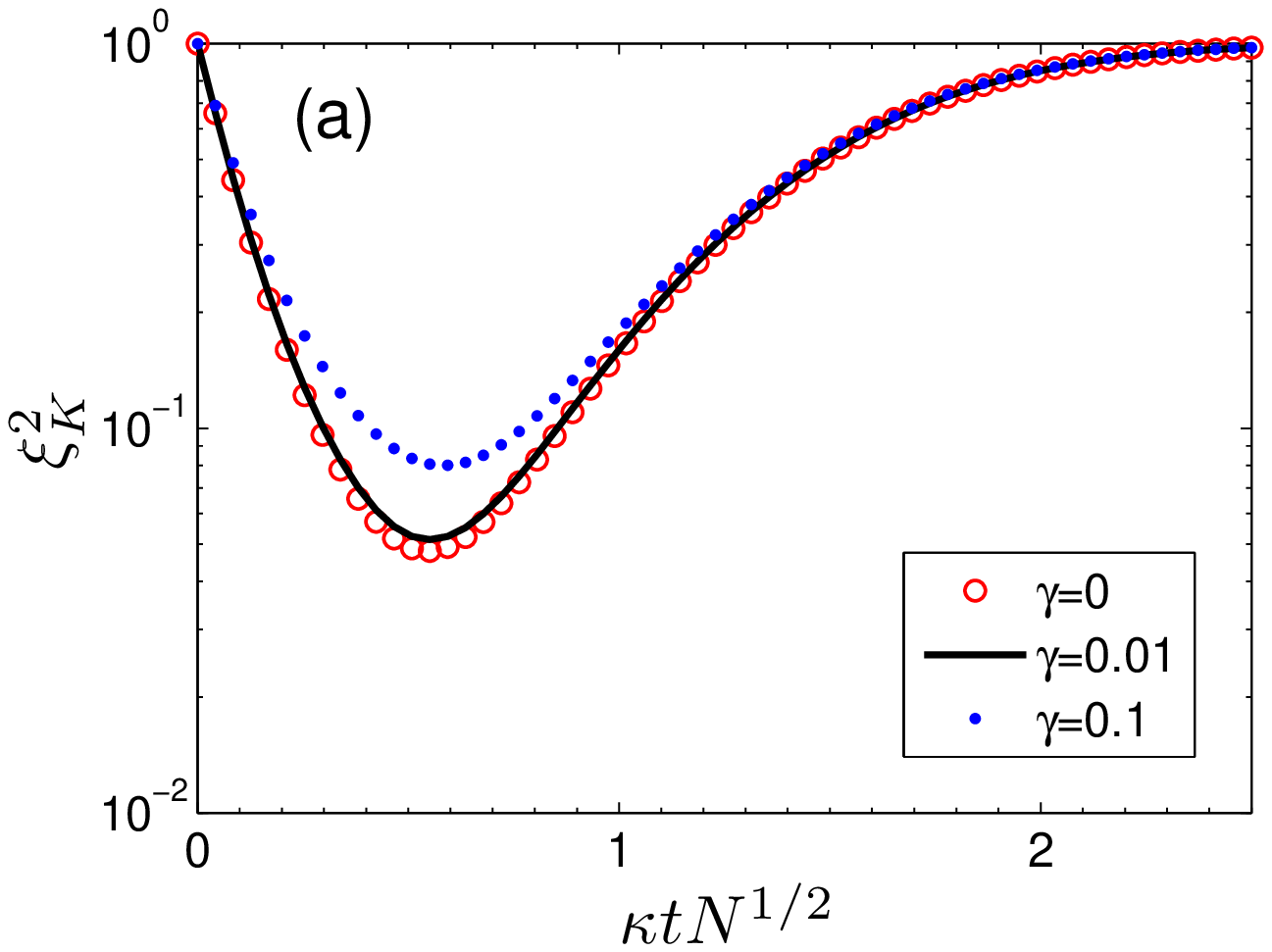}}
\subfigure{ \includegraphics[width=7.5cm,height=6cm]{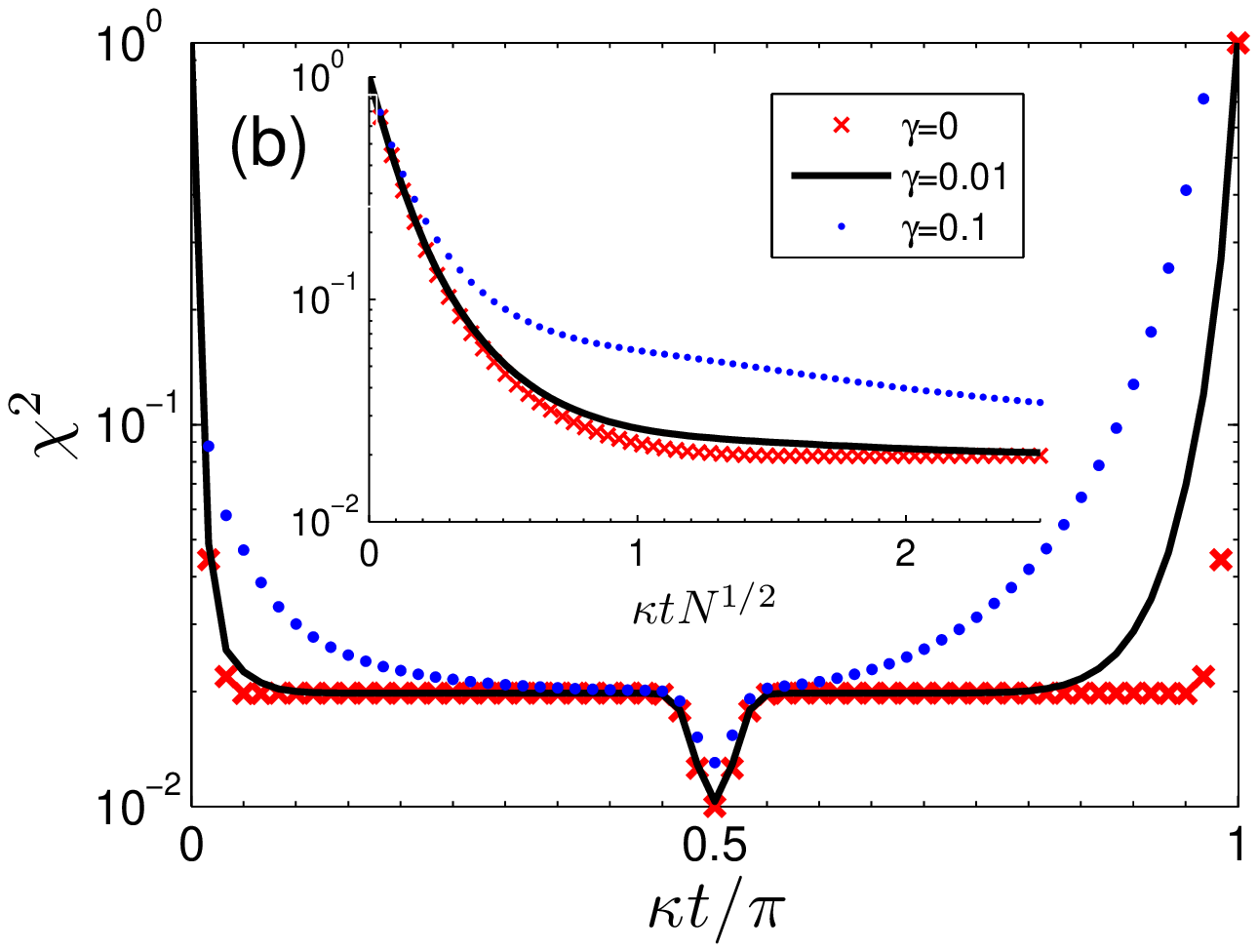}}
\caption{(Color online) Plots of the time evolution of $\xi_{K}^{2}$ (a) and $\chi^{2}$ (b) with $N=100$ for $\theta_{0}=\pi/2$ in terms of three cases: $\gamma=0$, $0.01$ and $0.1$. (b) Plots of $\chi^2$ for a period of time $\kappa T=\pi$ and inset is plotted for a short time range.}
\label{fig:fig2}
\end{figure}

In this case, $\xi_{K}^2$ can be exactly solved. Here, the state of system is denoted by the density matrix of equation (\ref{eq:master_eq result}). Then we find that $\langle\tilde{J}_{+}\rangle$, $\langle \tilde{J}_{+}(\tilde{J}_{z}+\frac{1}{2})\rangle$ and $\langle \tilde{J}_{+}^{2}\rangle$ equal to equations (\ref{eq:a1}), (\ref{eq:a2}) and (\ref{eq:a3}) by multiplying the factors of $e^{-\gamma \tau}$, $e^{-4\gamma \tau}$ and $e^{-\gamma \tau}$ respectively. Due to $[J^2,\mathcal{L}]=[J_z,\mathcal{L}]=0$, the expressions of $\langle\tilde{J}^{2}\rangle$, $\langle\tilde{J}_{z}\rangle$ and $\langle\tilde{J}_{z}^{2}\rangle$ remain unchanged as are given by equations (\ref{eq:J2}), (\ref{eq:Jz}), and (\ref{eq:J2z}). By submitting those solutions into equation (\ref{eq:V_pm}), one can exactly obtain $V_{-}$ and $\xi^2_{K}$. $\chi^2$ for the state of (\ref{eq:master_eq result}) can be numerically calculated based on equations (\ref{eq:QFI_mix}) and (\ref{eq:C_mix}).

In figure \ref{fig:fig2}(a) and (b), we plot the time evolutions of $\xi_{K}^2$ and $\chi^2$ in terms of different decay rates $\gamma=0,\,0.01,$ and $0.1$. As is shown in figure \ref{fig:fig2}(a), the degree of squeezing is weakened as $\gamma$ increases, while the time interval of squeezing remains unchanged for different values of $\gamma$. In figure \ref{fig:fig2}(b), it shows that the symmetry of $\chi^2$ is broken in the presence of collisional dephasing. With the strength of the decay rate increasing, the Heisenberg-scaling-limit precision can not be obtained. However, the plateau always exists even for the strong decay rate $\gamma=0.1$ only with the time at which $\chi^2$ arrives the plateau delayed. This means that one can still acquire the near-Heisenberg limit by extending the evolution time.

\section{Conclusion}
We have analytically studied the maximal QFI and spin squeezing in the OAT model with a CSS $\vert\theta_{0},\phi_{0}\rangle$. In \cite{Jin2009}, it has been found that spin squeezing depends sensitively on the polar angle $\theta_{0}$ of the initial CSS and the degree of spin squeezing degrades significantly when $\theta_{0}$ slightly deviates from the optimal angle $\pi/2$. Unlike $\xi_{K}^2$, the APEP $\chi^2$ is quite stable during the time evolution. In the time scales of $1/\sqrt{N}\leq\kappa t<\pi/2$ with  $N=2\times10^3$, $\chi^2$ always stays on the plateau level of $10^{-3}$ dB. Meanwhile, we found that $\chi^2$ is insensitive to the angle $\theta_{0}$ in the vicinity of $\pi/2$. Our results indicate that the OAT-induced state from a non-ideal initial CSS $\theta_{0}\sim\pi/2$ can still acquire a near-Heisenberg-scaling precision in parameter estimation. The common feature of $\xi_{K}^2$ and $\chi^2$ is that they are independent of the azimuth angle $\phi_{0}$ of the initial CSS.

Additionally, we considered the effects of the collisional dephasing on the dynamics of $\xi_{K}^2$ and $\chi^2$. The analytical expression of $\xi^2_K$ was obtained and $\chi^2$ was numerically calculated. With the strength of the decay rate $\gamma$ increasing, the degree of the maximal spin squeezing decrease obviously, however the time interval of squeezing remains unchanged. More importantly, our results show that by extending the evolution time $\chi^2$ can still reaches the near-Heisenberg limit even in the presence of collisional dephasing. Our work can have practical impact on precision estimation in quantum metrology with the OAT-induced state and can be implemented with BECs within current technology.

\ack
This work is supported by the NFRPC through Grant No. 2012CB921602, and the NSFC with Grants No. 11025527 and No. 10935010.

\appendix
\section{Calculation of expectations of the spin components}
In this appendix, we derive some formulas for calculating the expectations of the angular momentum. A coherent spin state $\left\vert \theta_{0},\phi_{0}\right\rangle $ can be expressed in the form of equation (\ref{eq:CSS}). The expectations of the spin components in table \ref{tab:table2} are calculated from
\begin{equation}
\langle\tilde{J}_{+}\rangle=j\sin\theta_{0}\,e^{i\phi_{0}}\left(\cos\frac{\mu}{2}+i\cos\theta_{0}\sin\frac{\mu}{2}\right)^{2j-1},
\label{eq:a1}
\end{equation}
which can readily be rewrite as
\begin{equation}
  \langle\tilde{J}_{+}\rangle=r\exp(i\phi),
\end{equation}
with the corresponding modulus and argument being
\begin{eqnarray}
  r &=& j\sin\theta_{0}(1-\sin^{2}\theta_{0}\sin^{2}\frac{\mu}{2})^{j-1/2},\\
  \phi &=& \phi_{0}+(2j-1)\arctan(\cos\theta_{0}\tan\frac{\mu}{2}).
\end{eqnarray}
Furthermore, we obtain
\begin{equation}
\langle \tilde{J}_{+}^{2}\rangle =j(j-\frac{1}{2})\,e^{2i\phi_{0}}\sin^{2}\theta_{0}\left(\cos\mu+i\cos\theta_{0}\sin\mu\right)^{2j-2}.
\label{eq:a2}
\end{equation}
Differentiating equation (\ref{eq:a1}) with respect to $\mu$ yields
\begin{equation}
\fl
 \langle \tilde{J}_{+}(\tilde{J}_{z}+\frac{1}{2})\rangle  = j(j-\frac{1}{2})\sin\theta_{0}e^{i\phi_{0}}\left(\cos\frac{\mu}{2}+i\cos\theta_{0}\sin\frac{\mu}{2}\right)^{2j-2}\left(i\sin\frac{\mu}{2}+\cos\theta_{0}\cos\frac{\mu}{2}\right).
 \label{eq:a3}
\end{equation}

\end{document}